\documentclass[epj]{svjour}
\usepackage{graphics}
\usepackage[latin1]{inputenc}
\usepackage{epsfig}
\usepackage{a4wide}
\usepackage{latexsym}
\usepackage{amssymb}
\usepackage{amsmath}
\usepackage{graphics}
\usepackage{cite}
\begin{document}
\def\d{{\rm d}}
\def\ex{{\rm e}}
\def\i{{\rm i}}
\def\rv{r_{\rm v}}
\def\e{{\bf e}}
\def\v{{\bf v}}
\def\u{{\bf u}}
\def\x{{\bf x}}
\def\r{{\bf r}}
\def\k{{\bf k}}
\def\l{{\bf l}}
\def\y{{\bf y}}
\def\pe{{a}}
\def\smalze{{\scriptscriptstyle (0)}}
\def\smalun{{\scriptscriptstyle (1)}}
\def\smaldu{{\scriptscriptstyle (2)}}
\def\smalk{{\scriptscriptstyle (k)}}
\def\bxi{{\boldsymbol{\xi}}}
\def\bnu{{\boldsymbol{\nu}}}
\def\beq{\begin{equation}}
\def\eeq{\end{equation}}
\title{Perturbation theory for large Stokes
number particles in random velocity fields}
\author{Piero Olla\inst{1} \and M. Raffaella Vuolo\inst{2}
}                     
%
%
\institute{ISAC-CNR and INFN, Sez. Cagliari, 
I--09042 Monserrato, Italy \and Dipartimento di Fisica and INFN, Universit\`a di Cagliari,
I--09042 Monserrato, Italy.}
\date{Received: date / Revised version: date}
%
\abstract{
We derive a perturbative approach to study, in the large inertia 
limit, the dynamics of solid particles in a smooth, incompressible 
and finite-time correlated random velocity field. We carry on 
an expansion in powers of the inverse square root
of the Stokes number, defined as the ratio of
the relaxation time for the particle velocities and
the correlation time of the velocity field.
We describe in  this limit the residual concentration fluctuations
of the particle suspension, and determine the contribution to the
collision velocity statistics produced by clustering.
For both concentration fluctuations and collision velocities,
we analyze the differences with the compressible one-dimensional
case.
} 
\authorrunning{Olla and Vuolo}
\titlerunning{Perturbation theory for large $S$ particles in random fields}
\maketitle
\section{Introduction}
\label{intro}
An important component in the transport of aerosols by turbulent flows is
the tendency to form clusters, an ubiquitous phenomenon that has been 
observed e.g.  inside clouds \cite{kostinski01}, and whose simplest 
instance is particle aggregation in a 1D (one dimensional)
random force field \cite{deutsch85}. 
In 1D, clustering is the combined result of particle
slipping with respect to the random field, and the fact that the random forces 
pushing particles apart become smaller as the separation decreases
\cite{deutsch85,wilkinson03}. In more than 1D, the picture is
more complicated, and the process is accompanied
by preferential concentration of the particles (supposed denser than
the fluid) in the strain 
regions of the flow \cite{wang93}. Thus, contrary to the intuition 
that would suggest a mixing behavior, a spatially 
homogeneous random velocity field
will lead to the formation of clumps out of an initially 
uniform distribution \cite{elperin96}. 
This unmixing will take the form of concentration fluctuations, 
superimposed to a mean concentration field that remains uniform,
and exeeding those due to discreteness in the distribution,
described by Poisson statistics
\cite{cencini06}. 
Additional segregation will be produced by 
inhomogeneity of the turbulent flow, such as in wall turbulence, 
resulting in a non-uniform mean concentration profile \cite{brooke94}.
Gravity will affect the concentration fluctuations of heavier particles 
\cite{fung03,vilela07}, decreasing
the effective correlation time of the turbulent fluctuations
sampled by the particles \cite{csanady63}.


A practical motivation for the interest in clustering is clearly the
possibility of enhanced binary collision, compared to a spatially 
homogeneous condition, and this could find application e.g. to rain 
formation \cite{shaw03,vaillancourt00}. It must be mentioned that there have been 
suggestions that clustering is of secondary importance in the collision 
dynamics \cite{falkovich02,wilkinson06}. It is not clear, in 
general, how clustering affects the relative velocity dynamics,
and in this way also collisions \cite{olla07}.

Several models for the clustering dynamics have been proposed
\cite{elperin96,sigurgeirsson02,zaichik03}, however,
although high Reynolds number turbulence is a multiscale flow, the bulk of
the studies has been on the dynamics of inertial particles in smooth random
fields (see e.g. \cite{balkovsky01,duncan05}). 
There are good reasons for this. The dynamics of a sufficiently
small (and sufficiently dense) spherical particle can be described by the
relaxation time of its velocity relative to the fluid: the Stokes time 
$\tau_S=2/9\ r_0^2\lambda/\nu_0$, where $r_0$ is the particle radius, 
$\lambda$ is the ratio of the particle to fluid density and $\nu_0$ is 
the kinematic viscosity of the fluid \cite{maxey83}. 
Now, most atmospheric aerosols
are characterized by Stokes times shorter than 
the Kolmogorov times of the flows by which they 
are advected \cite{shaw03}.
Furthermore, experimental data 
\cite{fessler94} and numerical simulations \cite{wang93} both indicate that 
clustering is stronger for particles with $\tau_S$ of the order of 
the Kolmogorov time, corresponding to a range of scales in which 
the velocity field could be approximated as smooth. 

In the case of a smooth incompressible random field, 
clustering is peaked at $S=\tau_S/\tau_E\sim 1$, where $\tau_E$ is the 
correlation time of the field and $S$ is called the Stokes number.
For an incompressible velocity field, the limit $S\to 0$ 
corresponds to passive scalar transport, and spatially
homogeneous initial conditions will not lead to clustering. In the
opposite limit $S\to\infty$, one expects
that the particles be scattered by the velocity
fluctuations they cross in their motion \cite{abrahamson75}
as if undergoing Brownian diffusion, resulting again
in no clustering. One reason for being interested in
this limit is that also particles in turbulent 
flows for which $\tau_S$ lies in the inertial range (or above), 
see smaller vortices as if $S\gg 1$. 
It must be said that the large $S$ limit is more of interest for industrial 
application, since in cloud turbulence, the relative motion of 
heavier droplets is dominated by the different gravitational 
settling velocity of particles of different size \cite{shaw03} and the
possibility of chaotic trapping \cite{stommel49,pasquero03}.

We shall focus in this paper on the case of a strongly stirred
fluid, such that, also for large-$S$ particles, the 
largest component of the particle-fluid
relative motion is due to inertia and not gravity.
We will try to understand, in particular, how the Brownian
limit of \cite{abrahamson75} is achieved.

There is some evidence \cite{bec07} that clustering destruction 
occurs at finite $S$, as the result of a crossover of the correlation 
dimension $D_2$ of the particle distribution, above the dimension of 
the space $D$.
For $S$ below this threshold, the probability
density function (PDF) $\rho(\r)$ of finding a pair of particles at separation
$\r$ diverges for $r\to 0$ like $r^{D_2-D}$ (see also
\cite{nishikawa01,balkovsky01}), while it should remain finite
above theshold. Actually, residual (non-singular) concentration fluctuations
remain present above threshold, decaying in 1D like
$\rho(\r=0)\propto S^{-1/2}$, and their origin lies in
the fact that only particles moving at increasingly 
small relative velocities, as $S\to\infty$, are able to stay close
long enough to remain correlated \cite{olla07}. 

The picture just described suggests the possibility of a perturbative expansion
around $S\to\infty$. This is formally identical to an expansion in powers of
the Peclet number $Pe$, for particle advection in the presence of strong molecular
diffusion. In both the large $S$ and the small $Pe$
regimes, the small quantity is the contribution to the relative particle motion
produced at small separation, by the correlations in the random field.
The perturbative approach will
allow to carry on an analysis of the
large-$S$ inertial particle dynamics for $D>1$,
where the heuristic approaches like the one in \cite{olla07} are made difficult
by the complexity of geometry and incompressibility.

This paper is organized as follows. In Sec. 2, the model equations
for the problem are derived. The perturbative approach is derived in 
Sec. 3.
In Sec. 4, the approach is tested in 1D,
comparing with numerical simulations and with the heuristic
results in \cite{olla07}. In Sec. 5, concentration fluctuations
are analyzed
in the case of a 3D incompressible random field. Section 6 is 
devoted to determination of the effect of concentration fluctuations
on the relative particle velocity statistics.
Section 7 contains the conclusions. 
Discussion of additional technical aspects is left to the appendices.

\section{The stochastic model}
\label{stochastic_model}
Following the derivation in \cite{olla07}, we introduce a
zero mean, smooth Gaussian random velocity field $\u(\x,t)$, 
with correlation
\beq
\langle u_\alpha(\r,t)u_\beta(0,0)\rangle=\sigma_u^2F(t)g_{\alpha\beta}(\r),
\label{eq1}
\eeq
where $g_{\alpha\beta}(0)=\delta_{\alpha\beta}$, 
$F(0)=1$, and we can introduce 
correlation times and lengths $\tau_E$ and $\rv$ obeying $\int_0^\infty\d\tau\, F(\tau)=\tau_E$,
and $\int_0^\infty\d r g_{\alpha\beta}(r\hat\r)\sim\rv$ ($\hat\r\equiv\r/r$ fixed). 
We assume isotropy and homogeneity of the field in space and time.

From the field parameters $\rv$, $\tau_E$ and $\sigma_u$, we can introduce the Kubo 
number $K=\sigma_u\tau_E/\rv$, which tells us whether the field is intrinsically
short- or long-time correlated. The $K\to 0$ limit would correspond to a random field
with zero correlation time, like in the Kraichnan model \cite{kraichnan94}. 
The $K\to\infty$ limit would correspond to a frozen field regime.
In most of our analysis we shall assume $K=O(1)$, as in actual turbulent flows. 

A suspension of inertial particles is advected by the random field
and their velocity $\v$ is taken to obey the Stokes equation:
\beq
\dot\v=\tau_S^{-1}[-\v(t)+\u(\x,t)]+\boldsymbol{\eta},\qquad\dot\x=\v,
\label{Stokes_equation}
\eeq
where we have allowed for a Brownian motion component $\boldsymbol{\eta}$:
$\langle\eta_\alpha(t)\eta_\beta(0)\rangle=\kappa\tau_S^{-2}\delta_{\alpha\beta}\delta(t)$; 
$\kappa$ is
the molecular diffusivity of the particles.

We choose units such  that $\sigma_u=\tau_S=1$; therefore: 
\beq 
\tau_E=S^{-1},
\qquad
r_\v=(KS)^{-1}.
\nonumber
\eeq
In the regime $S\gg 1$, the particle displacement in a correlation time $\tau_E$ becomes
negligible with respect to $\rv$ and Eq. (\ref{Stokes_equation}) could be approximated
by a Langevin equation (the precise condition for a Langevin dynamics 
is $K^2\ll S$ \cite{olla07}). 
It is then possible to substitute into Eq. (\ref{Stokes_equation}): 
\beq
\u(\x(t),t)\to (2\tau_E)^{1/2}\bxi(t),
\nonumber
\eeq
with $\bxi(t)$ white noise: 
$\langle\xi_\alpha(t)\xi_\beta(0)\rangle=\delta_{\alpha\beta}\delta(t)$, 
$\langle\xi_\alpha\eta_\beta\rangle=0$.
We thus obtain for the particle velocity variance 
$\sigma_v^2\sim S^{-1}(1+Pe^{-1})$, where
the Peclet number $Pe=2\sigma_u^2\tau_E/\kappa$ parameterizes the relative strength of
the random field and molecular contributions to particle diffusion. 

Turning to the relative motion of particle pairs, let us introduce
difference variables
$$
\bnu=\v_2-\v_1,
\qquad
\r=\x_2-\x_1,
$$
where 1 and 2 label members of a particle pair.  
(The particles are assumed immaterial, so that they can cross without interaction).
As in the one-particle case, 
we can approximate the equation for the relative motion of particles with the
Langevin equation:
\beq
\dot\nu_\alpha=-\nu_\alpha+b_{\alpha\beta}(\r)\xi_\beta,
\qquad
\dot r_\alpha=\nu_\alpha.
\label{Langevin2}
\eeq
where
\beq
b_{\alpha\gamma}(\r)b_{\gamma\beta}(\r)=
q^2[\delta_{\alpha\beta}-\pe g_{\alpha\beta}(\r)]
\label{B_alphabeta}
\eeq
(in this paper, summation over repeated indices is assumed), with 
\beq
\pe=(1+Pe)^{-1}Pe\quad{\rm and}\quad
q=2(S\pe)^{-1/2}.
\label{pe}
\eeq
We see that $\pe$ goes to zero with $Pe\to 0$, and, as expected, 
the relative particle dynamics
described in Eqs. (\ref{Langevin2}-\ref{B_alphabeta}) becomes uncorrelated.
The zero molecular diffusion limit corresponds instead 
to $Pe\to\infty$ and $\pe=1$.

It should be mentioned that, in the Langevin equation limit, the dynamics
becomes equivalent to that of a Kraichnan model and can be described in terms
of the single parameter \cite{wilkinson03}
\beq
\epsilon=K^2S.
\label{epsilon}
\eeq
To understand the connection with the Kraichnan model, notice that
$\epsilon^{-1}\tau_S\sim\rv^2/(\sigma_u^2\tau_E)$ 
is the diffusion time of a single passive tracer across a 
distance $\rv$ and plays the role of effective correlation time of the field,
so that $\epsilon$ becomes an effective Stokes number.
More interestingly, for $Pe,\epsilon\gg 1$, $\epsilon^{-1/2}\tau_S$ is the permanence time at
separation $<\rv=(KS)^{-1}$ of a particle pair approaching at relative
velocity $\sigma_v\sim S^{-1/2}$ (it is easy to see that such particles
cross at large enough speed to behave ballistically at scale $\rv$).


The limit $S\gg K\sim 1$, beyond allowing a Langevin equation based model,
leads to decoupling of the difference variables $(\bnu,\r)$ 
from the center of mass variables
$\frac{1}{2}(\x_1+\x_2)$ and $\frac{1}{2}(\v_1+\v_2)$. At equilibrium, 
this means that the two-particle PDF will be in the form, using 
spatial homogeneity:
$$
\rho(\v_{1,2},\x_{1,2})=\Omega^{-1}\rho
\Big(\frac{\v_1+\v_2}{2}\Big)\rho(\bnu,\r),
$$
where $\Omega^{-1}$, with $\Omega$ the volume of the domain for $\x$,  
is just $\rho((x_1+x_2)/2)$.  (Where not ambiguous,
we do not use subscripts to identify PDF's of different quantities).
Multiplying by $N^2$, with $N$ the total number of particles in
the domain and integrating over $\d^3v_1\d^3v_2$,
we obtain the expression for the concentration correlation:
\beq
\langle n(\r)n(0)\rangle=\bar n^2\Omega\int\d^3\nu\rho(\bnu,\r),
\label{eq9}
\eeq
where $\bar n=N/\Omega$ is the mean concentration. 
The quantity $f(r)=\Omega\rho(\r)-1$, with $\rho(\r)=
\int\d^3\nu\rho(\bnu,$ $\r)$,
gives the strength of the concentration fluctuations
(we have exploited isotropy). 
The concentration variance can be expressed in terms of 
the function $f$ by means of the relation
$\langle (n-\bar n)^2\rangle=\bar n^2f(0)$.

\section{Perturbation theory}
For small $Pe$, the two-particle PDF 
$\rho(\bnu,\r;t)=\langle\delta(\bnu(t)-\bnu)\delta(\r(t)-\r)\rangle$
can be determined solving Eqs. (\ref{Langevin2}-\ref{B_alphabeta}) 
for $\bnu(t)$ and $\r(t)$ perturbatively in $\pe$. 
We shall argue and verify numerically
in the next section, that the same perturbative strategy remains
valid for $S\gg K\sim 1$ also in the absence of molecular diffusion,
i.e. for $\pe=1$.

The perturbative expansion of $\bnu(t)$ and $\r(t)$ is obtained substituting the
Taylor expansion
$$
\begin{array}{ll}
b_{\alpha\beta}(\r(t))&=q[\delta_{\alpha\beta}-\frac{\pe}{2}g_{\alpha\beta}(\r(t))
\\
&-\frac{\pe^2}{8}g_{\alpha\gamma}(\r(t))g_{\gamma\beta}(\r(t))+\ldots],
\end{array}
$$
which is obtained from Eq. (\ref{B_alphabeta}),
into the solution of Eq. (\ref{Langevin2}):
$$
\begin{array}{ll}
r_\alpha(t)&=r_\alpha(-T)+\nu_\alpha(-T)(1-\ex^{-T-t})
\\
&+\int_{-T}^t\d\tau (1-\ex^{\tau-t})
b_{\alpha\beta}(\r(\tau))\xi_\beta(\tau),
\end{array}
$$
and Taylor expanding again in $\pe$.
The ground state of the expansion is therefore:
\beq
\begin{array}{ll}
b^\smalze_{\alpha\beta}&=q\delta_{\alpha\beta},
\\
r^\smalze_\alpha(t)&=r_\alpha(-T)+\nu_\alpha(-T)(1-\ex^{-T-t})
\\
&+q\int_{-T}^t\d\tau (1-\ex^{\tau-t}) \xi_\alpha(\tau),
\end{array}
\label{order0}
\eeq
while the first two corrections are:
\beq
\begin{array}{ll}
b^\smalun_{\alpha\beta}(t)=-(q\pe/2)g_{\alpha\beta}(\r^\smalze(t)),
\\
r^\smalun_\alpha(t)
=
\int_{-T}^t\d\tau (1-\ex^{\tau-t})
b^\smalun_{\alpha\beta}(\tau)\xi_\beta(\tau)
\end{array}
\label{order1}
\eeq
and 
\beq
\begin{array}{ll}
b^\smaldu_{\alpha\beta}(t)&=
-q[(\pe/2)r^\smalun_\gamma(t)\partial_\gamma g_{\alpha\beta}(\r^\smalze(t))
\\
&+(\pe^2/8)g_{\alpha\gamma}(\r^\smalze(t))g_{\gamma\beta}(\r^\smalze(t))],
\\
r^\smaldu_\alpha(t)&=\int_{-T}^t\d\tau (1-\ex^{\tau-t})
b^\smaldu_{\alpha\beta}(\tau)\xi_\beta(\tau),
\end{array}
\label{order2}
\eeq
where $\partial_\gamma\equiv\partial/\partial r_\gamma$.
We focus here on the PDF for the particle separation $\r$, leaving the analysis of
the joint PDF $\rho(\bnu,\r)$ to Sec. \ref{Collision}.
The PDF for $\r$, given some distribution of initial conditions $(\r(-T),\bnu(-T))$,
will be something in the form:
$$
\rho(\r;t)
=\int\prod_k\d^3 r^\smalk\rho(\{\r^\smalk\};t)
\delta(\sum_k\r^\smalk-\r).
$$
Depending on 
the choice of initial conditions for the trajectories ending at
$\r$, $\rho(\r;t)$ will be in general a non-equilibrium PDF. 
Carrying out the integral over $\r^\smalze$:
$$
\rho(\r;t)=\int\prod_k\d^3\tilde r^\smalk\rho(\{\tilde\r^\smalk\},\r^\smalze=\r-\tilde\r;t),
$$
where $\tilde\r^\smalk=\r^\smalk$ for $k>0$ and $\tilde\r=\sum_{k>0}\r^\smalk$.
Taylor expanding in $a$ leads to a perturbative series
$\rho=\rho^\smalze+\rho^\smalun+\ldots$, 
with $\rho^\smalze(\r,t)\equiv\rho(\r^\smalze(t)=\r)$. In explicit form:
$$
\begin{array}{ll}
\rho(\r,t)&=
[1-\partial_\alpha\langle r^\smalun_\alpha|\r\rangle
-\partial_\alpha\langle r^\smaldu_\alpha|\r\rangle
\\
&+(1/2)\partial_\alpha\partial_\beta\langle r^\smalun_\alpha r^\smalun_\beta|\r\rangle
+\ldots]
\rho^\smalze(\r,t)
\end{array}
$$
where
$\langle\r^\smalun|\r\rangle$ is a shorthand for
\beq
\Big\langle\r^\smalun(t)\Big|\r^\smalze(t)=\r;\ \r^\smalun(-T)=0\Big\rangle
\label{shorthand}
\eeq
and similarly for the other conditional averages. 
Sending $T\to\infty$ and using
$\lim_{T\to\infty}\rho(\r^\smalze)=\rho^\smalze(\r)=\Omega^{-1}$, we obtain the perturbative 
expansion for the equilibrium PDF:
\beq
\begin{array}{ll}
\Omega\rho(\r)=1&-\partial_\alpha\langle r^\smalun_\alpha|\r\rangle
-\partial_\alpha\langle r^\smaldu_\alpha|\r\rangle
\\
&+(1/2)\partial_\alpha\partial_\beta\langle r^\smalun_\alpha r^\smalun_\beta|\r\rangle
+\ldots
\end{array}
\label{perturb}
\eeq
Clearly, the dependence on the condition
$\r^\smalun($ $-T)=0$ in Eq. (\ref{shorthand})
disappears when $T\to\infty$, but the limit must be taken after
the average.

In order to take into account the simultaneous conditioning in the past and 
in the present in Eq. (\ref{shorthand}),
the following expression for $\langle\r^\smalun|\r\rangle$ can be utilized:
\beq
\Omega\Big\langle\r^\smalun(t)
\delta(\r^\smalze(t)-\r)\Big|\r^\smalun(-\infty)=0\Big\rangle,
\label{r1delta}
\eeq
where the factor $\Omega$ is simply
$[\rho(\r^\smalze(t)|r^\smalun(-\infty)$ $=0)]^{-1}$.
Substituting Eqs. (\ref{order0}) and 
(\ref{order1}) into (\ref{r1delta})
and setting without lack of generality $t=0$, we obtain:
$$
\Omega
\int_{-T}^0\d\tau(1-\ex^\tau)\langle b_{\alpha\beta}^\smalun(\tau)\xi_\beta(\tau)
\delta(\r^\smalze(0)-\r)\rangle.
$$
Using the functional integration by part formula \cite{zinn}, allows to treat
the correlation between $\xi_\beta(\tau)$ and the other factors in the integral:
$$
\int_{-T}^0\d \tau(1-\ex^\tau)\langle b_{\alpha\beta}^\smalun(\tau)
\frac{\delta r_\gamma^\smalze(0)}{\delta\xi_\beta(\tau)}\partial_{r_\gamma^\smalze(0)}
\delta(\r^\smalze(0)-\r)\rangle,
$$
with $\delta/\delta\xi_\beta(\tau)$ indicating functional derivative; from
Eq. (\ref{order0}): $\delta r_\gamma^\smalze(0)/\delta\xi_\beta(\tau)=
q(1-\ex^\tau)\delta_{\beta\gamma}$ [notice that
$\delta r^\smalze_\gamma(\tau)/\delta\xi_\beta(\tau)=0$]. Substituting 
$\partial_{r_\gamma^\smalze(0)}\to -\partial_\gamma$ and using Eq. (\ref{order1}),
we obtain finally:
\beq
\begin{array}{ll}
\langle r_\alpha^\smalun|\r\rangle&=
\frac{\Omega}{2}q^2\pe\partial_\beta\int_{-T}^0\d \tau(1-\ex^\tau)^2
\\
&\times
\langle g_{\alpha\beta}(\r^\smalze(\tau))\delta(\r^\smalze(0)-\r)\rangle.
\end{array}
\label{intermediate}
\eeq
Similar expressions can be derived for the higher orders in the expansion 
for $\rho$, and they will be required to take into account incompressibility
in the $D>1$ case.

\section{Small Peclet vs. large Stokes}
The two-particle distribution in Eq. (\ref{perturb}) is expressed as a formal (regular) 
perturbation expansion in $\pe$. The real expansion parameter, however, is
the ratio $\tilde r/\rv$, where $\tilde r\sim r^\smalun$ is the contribution 
of the velocity correlation $g_{\alpha\beta}$ to the relative particle displacement.
A relevant quantity is then the permanence time at $r^\smalze<\rv$, i.e.
the time during which $g_{\alpha\beta}(\r^\smalze)$ is significantly different from zero
and can contribute to the integral in Eq. (\ref{intermediate}). If $r<\rv$,
the permanence time will be $\rv/\nu\sim\rv/q$ [see Eq. (\ref{order0})].
Taking $\partial_\alpha,\partial_\beta\sim\rv^{-1}$ 
and $\delta(\r^\smalze-\r)\sim\Omega^{-1}$, gives therefore:
$$
\tilde r/r\sim\partial_\alpha\langle r^\smalun_\alpha|\r\rangle
\sim(\pe K/q)\,\min(1,(\rv/q)^2).
$$
Now, our perturbation expansion is carried on in powers
of $\pe$ for fixed $q$; in other words, $\pe$ (and therefore $Pe$) 
are considered small independently of $q$.
Had $K$ and $S$, in place of $q$, been chosen as the fixed quantities in our problem, 
we would have obtained instead, from Eq. (\ref{pe}):
\beq
\tilde r/\rv
\sim\pe^{3/2}\epsilon^{1/2}\,\min(1,\pe/\epsilon).
\label{small}
\eeq
This suggests that the perturbation expansion of Eq. (\ref{perturb}) could be translated, 
for large $\epsilon$ and $Pe$ generic, 
into a new expansion in powers of
$\epsilon^{-1/2}$. 

The validity of the new expansion rests on regularity of the solution $\rho(\r)$
in $\r=0$ (the estimate $\partial_\alpha,\partial_\beta\sim \rv^{-1}$  
adopted implicitly this assumption). While this is guaranteed for small $\pe$ 
by molecular diffusion, for $\pe=1$, $b_{\alpha\beta}(0)=0$ and Eq. (\ref{Langevin2})
becomes singular. 
Now, the correspondence between the 
small $\pe$ and large $\epsilon$ cases guarantees finiteness of the
terms in the series (it is possible to see that this is not verified 
in the case a ground state of ballistic particles at scale $\rv$ is
adopted \cite{olla07}). Hence, expanding in powers of $\epsilon^{-1/2}$
is a meaningful procedure.
It is still possible that the perturbative approach is simply unable
to catch singular behaviors [notice that Eq. (\ref{small}) does not reveal 
break-down of the perturbation theory for $\pe=1$ and $\epsilon\ll 1$, 
in spite of the singularity of $\rho(\r)$]. The numerical evidence in 
\cite{bec07}, that singular behavior are absent 
for large $\epsilon$, however, suggests that this is not the case.

Let us test in 1D the perturbative approach just introduced.
For the sake of definiteness, we consider a Gaussian profile for 
$\langle u(r,t)u(0,t)\rangle$: 
\beq
g(r)=\exp(-r^2/2\rv^2);
\label{g(r)}
\eeq
hence: $b^\smalze=q$ and $b^\smalun=(q\pe/2)\, g(r^\smalze)$.

Substituting Eq. 
(\ref{g(r)}) into 
(\ref{intermediate})  and then into
(\ref{perturb}), we can write $\rho^\smalun(r)$ in the 
compact form:
\beq
\rho^\smalun(r)=-\frac{q^2a}{2}\int_{-T}^0\d t H_r(t)
A_r(r,t),
\label{rho1}
\eeq
with 
$A_r(r,t)=\int\d r^\smalze(t)\rho(r^\smalze(t),r^\smalze(0)$=$r)g($ $r^\smalze(t))$
and
$H_r(t)=(1-\ex^t)^2\partial_r^2$;
the averages have eliminated all dependence on the initial conditions at $t=-T$.
We can write $\rho(r^\smalze(t),r^\smalze($ $0))=\Omega^{-1}\rho(r^\smalze(t)|r^\smalze(0))
=\Omega^{-1}\rho(s(t))$, with $s(t)=r^\smalze(t)-r^\smalze(0)$, 
and $\rho(s(t))$, for $T\to\infty$, is a zero mean Gaussian 
with variance, from Eq. (\ref{order0}): 
\beq
\sigma^2_{s(t)}=q^2[-1-t+\ex^t].
\label{sigma2s}
\eeq
Substituting into Eq. (\ref{rho1}), we obtain the result
\beq
\begin{array}{ll}
f(r)&=\frac{\pe^{5/2}}{4\epsilon^{1/2}}\int_0^\infty
\d t\frac{(1-\ex^{-t})^2}{G_r^{3/2}(\epsilon,t)}
\\
&\times\Big[1-\frac{\bar r^2}{G_r(\epsilon,t)}\Big]
\exp\Big\{-\frac{\bar r^2}{2G_r(\epsilon,t)}\Big\};
\\
\\
G_r(\epsilon&,t)=t-1+\ex^{-t}+\frac{\pe}{4\epsilon},
\end{array}
\label{f(r)}
\eeq
where $\bar r=r/q$. Notice that $q$ is the typical separation of a pair
of particles at a time $\sim\tau_S=1$ after crossing ($\bar r$ played the role of  
outer scale in the matching asymptotics analysis in \cite{olla07}).
In other words, the correlation length of the concentration fluctuations
is not expected to be of the order of $\rv$, rather, of the larger
scale $q\sim(\epsilon/\pe)^{1/2}\rv$.

\subsection{The 1D case: small $\epsilon$ regime}
Taking the limit $\epsilon\to 0$ under the condition $K^2/S\ll 1$, 
guarantees that the Langevin approach of Sec. \ref{stochastic_model} continues to
be valid. (Notice that the above conditions imply $K\ll 1$).
This regime has been extensively studied in \cite{wilkinson03,duncan05}
as a stochastic model for the behavior of low inertia $S\ll 1$ particles 
in realistic turbulent flows, for which $K\sim 1$.

The $\epsilon\to 0$ limit is diffusive at scale $\rv$ for the variable
$r(t)$ and we can derive an exact expression for $\rho(r)$, 
valid for all $\pe<1$. 
This expression can then be used
to test the perturbative results of the previous section. 
A diffusive limit means that the variable $r(t)$ changes in a correlation 
time $\tau_S$ much less than the spatial scale $r_\v$ [on the contrary, for
most pairs, the
large $\epsilon$ limit is ballistic at scale $\rv$ and becomes
diffusive only at the scale $(\epsilon/a)^{1/2}\rv$].
The drift and diffusivity in the effective equation for $r$ are the
leading contributions to
$$
\langle [r(t)-r(0)]|r(0)\rangle/t
\ \ 
{\rm and}
\ \ 
\langle [r(t)-r(0)]^2|r(0)\rangle/t,
$$
for $t\gg 1$ and $|r(t)-r(0)|\ll r_\v$. 
Actually, it is easy to prove from moment analysis of the Fokker-Planck equation 
for Eqs. (\ref{Langevin2}),
that the drift $\langle \dot r|r\rangle=\langle\nu|r\rangle$ 
is zero \cite{olla07}. We only have to calculate the diffusivity; from Eq. (\ref{Langevin2}), we
can write, for $t\gg 1$ and $|r(t)-r(0)|\ll r_\v$:
$$
r(t)=r(0)+b(r(0))\int_0^t\d\tau\xi(\tau)+O(t^0)
$$
and $\langle [r(t)-r(0)]^2|r(0)\rangle/t\simeq B(r(0))$. 
The particle separation behaves therefore like a non-uniform Brownian motion:
$\dot r=b(r)\xi$ and from the associated Fokker-Planck equation we obtain 
the exact expression:
\beq
f(r)=[1-\pe\exp(-\frac{1}{2}(KSr)^2)]^{-1}.
\label{small-epsilon}
\eeq
Notice the quadratic divergence of the PDF at $r\to 0$ for $\pe=1$, 
that is approached at $t\to\infty$ with an $O(\epsilon)$ rate \cite{wilkinson03}.

Let us compare with Eq. (\ref{f(r)}). In the limit $\epsilon\to 0$, we
can write $G_r(\epsilon,t)=t+\pe/(4\epsilon)$ and Eq. (\ref{f(r)}) becomes, 
after a few manipulations:
$$
f(r)=
\pe\exp(-\frac{1}{2}(KSr)^2),
$$
but this is precisely the $O(\pe)$ contribution to Eq. (\ref{small-epsilon}).

\subsection{The 1D case: large $\epsilon$ regime}
The prediction in \cite{olla07} that $f(0)\sim\epsilon^{-1/2}$ for $\epsilon\to\infty$
seems to be supported by Eqs. (\ref{small}) and (\ref{f(r)}).
However, the integral in Eq. (\ref{f(r)})
presents singularities for $\epsilon\to\infty$, and we must put some care in the
analysis.

Let us consider first the case $r=0$, and write Eq. (\ref{f(r)}) in the
form
\beq
f(0)=\frac{\pe^{5/2}}{\sqrt{2\epsilon}}
\Big[\int_0^T\frac{t^2\d t}{(t^2+\pe/(2\epsilon))^{3/2}}+\ldots\Big],
\nonumber
\eeq
where $\epsilon^{-1/2}\ll T\ll 1$; the dots indicate the remnant of the integral in 
Eq. (\ref{f(r)}), which is shown by inspection to be finite in the limit $\epsilon\to\infty$.
The first integral in the formula above, instead, is dominated by the ballistic 
crossing time
$t\sim\epsilon^{-1/2}$ and is logarithmically divergent for $\epsilon\to\infty$. We
obtain the leading order expression for $f(0)$:
\beq
f(0)=\frac{\pe^{5/2}\ln\epsilon}{2\sqrt{2\epsilon}}
+\ldots
\nonumber
\eeq
The logarithmic divergence is eliminated and a pure power law is recovered, 
provided we coarse grain the function $f$ at a fixed scale $R$:
$$
f_R(r)=(1/R)\int w(r'/R)f(r-r')\d r',
$$
where $w(r)$ is a smoothing function with $w(r)>0$ and $\int w(r)\d r=1$. 
A simple
analytic expression for $f_R$ is obtained choosing 
a Gaussian $w(r)$ $=(2\pi)^{-1/2}\exp(-r^2/2)$:
\beq
f_R(0)=\frac{\pe^{5/2}}{4\sqrt{\epsilon}}
\int_0^\infty\frac{(1-\ex^{-t})^2\d t}{[t-1+\ex^{-t}+\bar R^2]^{3/2}},
\label{coarse_grained}
\eeq
where $\bar R=R/q$.

\begin{figure}
\begin{center}
\includegraphics[draft=false,width=6.5cm]{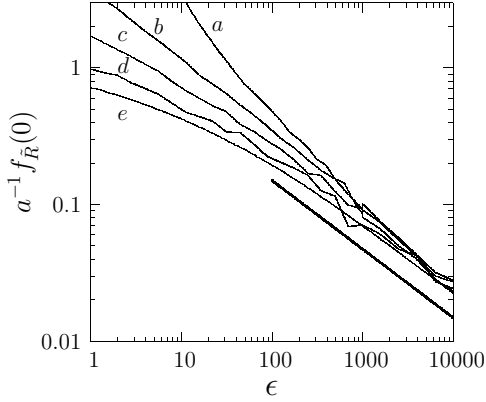}
\caption{
Rescaled fluctuation amplitude profiles for different values
of $\pe$, coarse grained at scale $\bar R=0.08$.
Cases $a-d$ are from from numerical integration of Eqs. 
(\ref{Langevin2}-\ref{B_alphabeta}): ($a$) $\pe=1.0$; ($b$) $\pe=0.9$; ($c$) $\pe=0.7$;
($d$) $\pe=0.4$. Case ($e$) is the theoretical prediction from Eq. (\ref{coarse_grained}).
The heavy line is $\epsilon^{-1/2}$.
}
\label{pertfig1}
\end{center}
\end{figure}

These results confirm the heuristic 
prediction in \cite{olla07}. The interesting point is the
improved performance of perturbation theory at large $\epsilon$, as
illustrated in Fig. \ref{pertfig1}. The collapse of the rescaled profiles indicates 
a region in which the higher orders in the perturbation expansion 
are playing a negligible role. 
%
%
The perturbation expansion in $\pe$ can therefore be converted into one in 
$\epsilon^{-1/2}$, as claimed.

We focus next on the correlation profile $f(r)$. We can obtain analytical 
expressions for large $\bar r=r/q$. We rewrite Eq. (\ref{f(r)}) in 
the form
\beq
f(r)=\frac{\pe^{5/2}}{4\sqrt{\epsilon}}
\partial_{\bar r}\int_0^\infty \bar r^4h^2(x)U(x+h(x))\d x,
\label{intermediate1}
\eeq
where $x=t/\bar r^2$, $h=\bar r^{-2}[\exp(-\bar r^2x)-1]$ and
$U(x)=x^{-3/2}\exp(-1/(2x))$. We can Taylor expand $U(x+h(x))=
U(x)+h(x)U'(x)+...$. We then substitute into Eq. (\ref{intermediate1}) and integrate 
by parts, using the  
fact that $R$ and all its derivatives are zero at both zero and infinity. 
The integrand in Eq. (\ref{intermediate1}) takes the form:
\begin{eqnarray}
\ r^4[h^2-(h^3)'+\frac{1}{2}(h^4)''-...]U(x)
\nonumber
\\
= [1+a_1\ex^{-\bar r^2x}+a_2\ex^{-2\tilde r^2x}+...]U(x).
\label{intermediate2}
\end{eqnarray}
We have $a_1=-\frac{2}{0!}+\frac{3}{1!}-\frac{4}{2!}+...=-\exp(-1)$
and the leading order in Eq. (\ref{intermediate2}) is therefore
$[1-\exp(-1-\tilde r^2 x)]x^{-3/2}\exp(-1/(2x))$. Substituting into Eq.
(\ref{intermediate1}), we get the final expression:
\beq
f(r)=\frac{\pi\pe^{5/2}}{2\sqrt{\epsilon}}\exp(-\sqrt{2}|\bar r|-1)+O(\ex^{-2|\bar r|}).
\label{profile}
\eeq
Notice that, opposite to the case of $f(0)$, the integral in Eq. (\ref{intermediate1})
is dominated by the long time scale $t\sim\bar r$, corresponding to $x\sim\bar r^{-1}$.
The heuristic predictions on the correlation profile in \cite{olla07} are therefore confirmed.
As shown in Fig. 2, however, agreement with the scaling prediction of Eq. (\ref{profile}) 
is obtained only for rather large values of $\epsilon$ and of the rescaled 
separation $\bar r$.

%
%
\begin{figure}
\begin{center}
\includegraphics[draft=false,width=7cm]{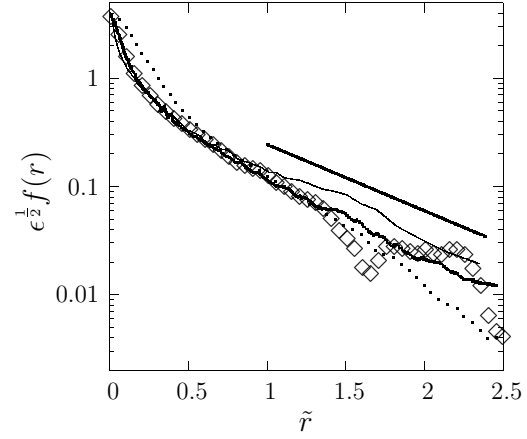}
\caption{
Fluctuation correlation profiles for $\pe=0.9$ and 
three different values of $\epsilon$. 
Dotted line: $\epsilon=10$; heavy line: $\epsilon=100$; thin line:
$\epsilon=1000$.  Diamonds correspond again to the 
$\epsilon=100,\pe=0.9$ case, but from direct simulation of 
Eq. (\ref{Stokes_equation}), instead of Eqs. (\ref{Langevin2}-\ref{B_alphabeta}).
Simulation parameters in the case of the full simulation:  
$K=1$, $N=10^4$ particles, total simulation time $=100\tau_S$;
domain length $\Omega=10S^{-1/2}$, corresponding to $512$ modes in 
the Fourier decomposition of the random field $u(x,t)$. 
The straight line is the $f(r)\propto\exp(-\sqrt{2}\bar r)$ prediction
of Eq. (\ref{profile}).
}
\label{pertfig2}
\end{center}
\end{figure}

\section{Concentration fluctuations in 3D}
From now on we restrict the analysis to the regime $Pe\to\infty$ (i.e. $\pe=1$). This is
physically consistent with a large $S$ regime, corresponding to particles with 
high inertia in the presence of strong advection. We recall that for $K\sim 1$, 
$S\gg 1$ corresponds to $\epsilon\gg 1$.
In more than 1D, if the flow is incompressible, concentration fluctuations will appear
only at second order in the perturbative expansion of Eq. (\ref{perturb}). In fact,
the generalization to more than 1D of Eq. (\ref{rho1}) reads:
$$
\begin{array}{ll}
\rho^\smalun(\r)=&-(2/S)\int_{-T}^0\d t(1-\ex^t)^2
\\
&\times
\partial_\alpha\partial_\beta
\int\d^3r^\smalze(t)g_{\alpha\beta}(\r^\smalze(t))
\\
&\times
\rho[\r^\smalze(t))|\r^\smalze(0)=\r],
\end{array}
$$
that will be identically zero if $\partial_\alpha g_{\alpha\beta}=0$. 
The lowest order contribution to $f(r)$ is therefore:
\beq
\Omega\rho^\smaldu(\r)=
-\partial_\alpha\langle r^\smaldu_\alpha|\r\rangle
+\frac{1}{2}\partial_\alpha\partial_\beta\langle r^\smalun_\alpha r^\smalun_\beta|\r\rangle,
\nonumber
\eeq
where $\r^\smalun$ and $\r^\smaldu$ are given in Eqs. (\ref{order1}-\ref{order2}). 
Contrary to the heuristic prediction in \cite{olla07}, of an $\epsilon^{-1/2}$
decay at large $\epsilon$ for the concentration fluctuation variance, we thus
expect  an $\epsilon^{-1}$ behavior. We show that this estimate is correct in 
Appendix A,  by explicit
calculation of $\rho^\smaldu$.

\section{Relative velocity statistics}
\label{Collision}
The approach in Sec. III can be extended to the calculation of the joint
PDF $\rho(\r,\bnu)$, with Eqs. (\ref{perturb},\ref{r1delta}) 
being replaced by
\beq
\begin{array}{ll}
\rho(\y)&=
[1-\partial_\alpha\langle y^\smalun_\alpha|\y\rangle
-\partial_\alpha\langle y^\smaldu_\alpha|\y\rangle
\\
&+\frac{1}{2}\partial_\alpha\partial_\beta\langle y^\smalun_\alpha y^\smalun_\beta|\y\rangle
+\ldots]
\rho^\smalze(\y)
\end{array}
\label{perturby}
\eeq
and
$$
\begin{array}{ll}
&\langle\y^\smalun|\y\rangle\rho^\smalze(\y)
\\
&=\Big\langle\y^\smalun(t)
\delta(\y^\smalze(t)-\y)\Big|\y^\smalun(-T)=0\Big\rangle,
\end{array}
$$
where $\y=(\r,\bnu)$ and the vector indices run now from 1 to 6.
Clearly, integrating Eq. (\ref{perturby}) over $\bnu$ leads to Eq. (\ref{perturb}).

From Eq. (\ref{perturby}), we could in principle obtain information on the velocity
components that contribute the most to clustering.
Looking at Eq. (\ref{perturby}), we see that the terms contributing to $\rho^\smalk(\r)$
for $k>0$ are those containing only spatial derivatives (the others give zero
after integration over $\bnu$). We can thus write:
\beq
\rho(\r,\bnu)=[\rho^\smalze(\bnu)+\rho_{cl}(\bnu|\r)]\rho(\r)+\Omega^{-1}\rho_B(\bnu|\r),
\label{rho_cl}
\eeq
where
$\Omega^{-1}\rho_B(\bnu|\r)$ contains the terms in the RHS of Eq. (\ref{perturby})
involving velocities derivatives $\partial_{\nu_\alpha}
\equiv\partial/\partial y_{3+\alpha}$ that do not contribute to Eq. 
(\ref{perturb}); 
of course, 
$\int\d^3\nu\,\rho_B(\bnu|\r)=\int\d^3\nu\,\rho_{cl}(\bnu$ $|\r)=0$.
Substituting into Eq.
(\ref{perturby}):
\beq
\begin{array}{ll}
\rho_{cl}(\bnu|\r)&=\Big\{[\Omega\rho(\r)]^{-1}
\Big[1-\partial_\alpha\langle r^\smalun_\alpha|\y\rangle
\\
&-\partial_\alpha\langle r^\smaldu_\alpha|\y\rangle
+\frac{1}{2}\partial_\alpha\partial_\beta\langle r^\smalun_\alpha r^\smalun_\beta|\y\rangle
\\
&+\ldots\Big]-1\Big\}\,
\rho^\smalze(\bnu).
\end{array}
\label{intermediate4}
\eeq

We may thus interpret $\rho_{cl}(\bnu|\r)$ as the cluster contribution to 
the velocity PDF $\rho(\bnu|\r)$ at separation $\r$. 
It is important to notice that this contribution is not necessarily localized in the
clusters: different moments with respect to $\bnu$ of $\rho_{cl}(\bnu,\r)=
\rho_{cl}(\bnu|\r)\rho(\r)$ do not come necessarily from the same
spatial spots in the volume $\Omega$.

Contrary to Eq. (\ref{perturb}), the expansion parameter in the
velocity part of Eq.  (\ref{perturby}) is now velocity dependent.
In analogy with Eq. (\ref{intermediate}), we can write 
\beq
\begin{array}{ll}
\rho_{cl}^\smalun(\bnu,\r)&=
\frac{\Omega}{2}q^2\partial_\alpha\partial_\beta\int_{-T}^0\d \tau(1-\ex^\tau)^2
\\
&\times
\langle g_{\alpha\beta}(\r^\smalze(\tau))\delta(\y^\smalze(0)-\y)\rangle.
\\
&\sim q^2\rv^{-2}\tau^3_{exit}(\bnu|\r)\rho^\smalze(\bnu)
\end{array}
\label{smallv}
\eeq
where $\tau_{exit}(\bnu|\r)$ is the permanence time at separation $<\rv$ of a particle pair,
characterized at given time by values $(\r,\bnu)$ of the relative position and velocity; 
assuming ballistic motion and taking $r=0$: $\tau_{exit}\sim\rv/\nu$. 
For typical particle pairs, for
which $\nu\sim\sigma_v\sim S^{-1/2}$, we thus have 
$\rho^\smalun\sim\epsilon^{-1/2}\rho^\smalze$. 
For $\nu\sim\epsilon^{-1/6}\sigma_v$, though, we find
$\rho^\smalun\sim\rho^\smalze$ and perturbation theory breaks down.

To understand what happens, we derive from Eq. (\ref{Langevin2}) the analog
of the lowest order equation for the position (\ref{order0}), in the case of
the velocity: 
\beq
\nu^\smalze_\alpha(t)=\nu_\alpha(-T)\ex^{-t-T}+q\int_{-T}^t\d\tau\ex^{\tau-t}\xi_\alpha(\tau).
\label{order0v}
\eeq
Integrating Eq. (\ref{order0}) with initial conditions $(\bnu=0,\r=0)$, we see that
the particles separate at $r^\smalze=\rv$ at a time
$\sim\epsilon^{-1/3}\tau_S$, which is the lowest order estimate for $\tau_{exit}(0|0)$. From 
Eq. (\ref{order0v}), $\epsilon^{-1/6}\sigma_v$ appears to be the relative
velocity at $t=\tau_{exit}(0|0)$, and is therefore the escape velocity out of the
correlated region $r\lesssim\rv$ for slow particle pairs, that do not behave ballistically
in that region \cite{olla07}. 
Notice that there are no singularities at $\nu=0$, and, 
in contrast with a perturbation 
theory with a ground state of ballistic particles, the individual terms in 
the expansion remain finite as $\nu\to 0$.

Here, we focus on the contribution to the collision velocity variance
$\langle\nu^2|\r$=$0\rangle_{cl}$ $=\int\d^3\nu\,\nu^2\rho_{cl}($ $\bnu|\r=0)$.
Contrary to perturbation theory for $\rho(\bnu)$, that breaks up for
$\nu\sim \epsilon^{-1/6}\sigma_v$, the one for the moments of $\rho$ is
perfectly well behaved, as
the contribution from $\nu<\epsilon^{-1/6}$ to
$\langle\nu^p|\r\rangle_{cl}=\int\d^3\nu\,\nu^p\rho_{cl}(\bnu|\r)$
is $\sim\epsilon^{-(3+p)/6}\rho_{cl}(0|\r)$, and $\rho_{cl}(0$ $|\r)$
is finite.

Substituting into Eq. (\ref{intermediate4}), we obtain
\beq
\begin{array}{ll}
\langle\nu^2|\r\rangle_{cl}&=
-\partial_\alpha\langle r^\smalun_\alpha(\nu^\smalze)^2|\r\rangle
-\partial_\alpha\langle r^\smaldu_\alpha(\nu^\smalze)^2|\r\rangle
\\
&+\frac{1}{2}\partial_\beta\partial_\gamma
\langle r^\smalun_\beta r^\smalun_\gamma(\nu^\smalze)^2|\r\rangle+\ldots
\\
&-f(r)\langle (\nu^\smalze)^2\rangle
+f(r)\partial_\alpha\langle r^\smalun_\alpha(\nu^\smalze)^2|\r\rangle
\\
&+f^2(r)\langle (\nu^\smalze)^2\rangle+\ldots
\end{array}
\label{<nu^2>_2}
\eeq
and we recall that $f(r)=\Omega\rho(r)-1$ [see Eq. (\ref{eq9})]. 
Again, all conditional averages are intended in the sense of Eq. (\ref{shorthand}).
Analogously to the analysis in Secs. IV and V, we see that the first non-zero contribution 
in the incompressible case arises at second order:
\beq
\begin{array}{ll}
\langle\nu^2|\r\rangle^\smaldu_{cl}=
&-\partial_\alpha\langle r^\smaldu_\alpha(\nu^\smalze)^2|\r\rangle
-\Omega\rho^\smaldu(r)\langle (\nu^\smalze)^2\rangle,
\\
&+\frac{1}{2}\partial_\beta\partial_\gamma
\langle r^\smalun_\beta r^\smalun_\gamma(\nu^\smalze)^2|\r\rangle,
\\
\end{array}
\label{nu23D}
\eeq
while in the 1D case:
\beq
\langle\nu^2|r\rangle^\smalun_{cl}=
-\partial_r\langle r^\smalun (\nu^\smalze)^2|r\rangle
-\Omega\rho^\smalun(r)\langle (\nu^\smalze)^2\rangle.
\label{nu21D}
\eeq

The calculation of the cluster contribution to the collision velocity variance
at this point is a matter of lengthy but straightforward algebra.
Leaving the calculation details to 
Appendix B, the result in 1D and in the incompressible 3D case 
are [see Eqs. (\ref{finalA},\ref{finalB})]: 
\beq
\langle(\nu^\smalze)^2\rangle^{-1}\langle\nu^2|0\rangle_{cl}^\smalun
=-4(\pi\epsilon)^{-1/2}\ln cS,
\label{nu21Dfinal}
\eeq
and
\beq
\langle(\nu^\smalze)^2\rangle^{-1}
\langle\nu^2|r=0\rangle^\smaldu_{cl}
=H\epsilon^{-1/2},
\label{nu23Dfinal}
\eeq
with $c$ and $H$ positive $O(1)$ constants. 

We may interpret these results by saying that concentration fluctuation produce 
collision hindering in the compressible 1D case, and to 
collision enhancement in the incompressible 3D case.

Notice the $O(\epsilon^{-1/2})$ behavior of $\langle\nu^2|r=0\rangle^\smaldu_{cl}$
in place of the expected $O(\epsilon^{-1})$ at second order. It is possible to show,
however,
that the $O(\epsilon^{-1/2})$ contributions from $\rho_{cl}$ and $\rho_B$ cancel, 
so that the total correction to the collision velocity variance is $O(\epsilon^{-1})$.

\section{Conclusion}
We have derived a perturbative approach for the two-particle statistics of
a randomly advected inertial particle suspension, that is valid 
both in regimes
of small Peclet number $Pe$ (strong molecular diffusion), and large Stokes
number $S$ (corresponding to high inertia). In both cases, one expands around 
a lowest order of independent Brownian particles. This is natural in the small 
$Pe$ regime; for large $S$ [more precisely, for large $\epsilon$, with
$\epsilon$ defined in Eq. (\ref{epsilon})], the mechanism is more subtle
and is due to the fact that
particle trajectories evolve on 
a characteristic scale $\epsilon^{1/2}\rv$, much larger
than the correlation length $\rv$ of the field $\u$. The
particle pair trajectories are dominated therefore by 
large separation, uncorrelated $\u$ contributions \cite{abrahamson75}.

The perturbative approach we have derived provides an analytical description
of how the Brownian particle limit of \cite{abrahamson75} is achieved, 
that goes beyond the qualitative considerations in \cite{olla07}.
The concentration fluctuation
amplitude $\bar n^{-2}\langle (n-\bar n)^2\rangle$, with $\bar n$ the
mean concentration, appears to be $O(\epsilon^{-1/2})$ for compressible,
and $O(\epsilon^{-1})$, for incompressible flows. The correlation length
of the fluctuations is $\sim\epsilon^{1/2}\rv$, in contrast with the
$\epsilon\ll 1$ case, in which, a power law at separations $r<\rv$ would
occur.

The perturbative approach allows to identify a concentration fluctuation 
contribution to the statistics for the collision velocity $\bnu$, 
in those terms in the expansion for the
joint PDF $\rho(\bnu,\r=0)$, that lead to deviations 
from the uniform fluctuation-free regime in the separation PDF $\rho(\r)$.
This goes beyond the observation that the expected relative velocity 
should decrease at small values of the separation $\r$.

It has been noted in \cite{olla08} that for $\epsilon\ll 1$, 
different collision velocities originate
from particle ''jumps'' starting at different initial separations. 
The collision velocity distribution is thus affected by
the spatial structure of the clusters, which produce an effective collision 
hindering.

In the present large $\epsilon$ regime, no such direct physical 
association between clustering and collision dynamics exists.
From an analysis of the collision velocity variance, 
we see that collision hindering occurs in 1D, while enhancement occurs in 3D 
if the flow is incompressible.  The result in 1D is not
unexpected, as clusters are in this case the result of particles slowing
down relative to one another as they get closer. The role of incompressibility
and 3D in leading to collision enhancement is less clear.

Additional information on the velocity statistics is contained in 
the velocity structure of the concentration fluctuations.
From  Eq. (\ref{smallv}), we see that, for $D<3$, the integral
$\rho(\r)=\int\d^D\bnu\rho($ $\bnu,\r)$ is dominated by small relative velocities,
while for $D=3$ all velocities 
contribute equally, down to the velocity scale $\epsilon^{-1/6}\sigma_v$,
at which, particle pairs with $r<\rv$ cease to behave ballistically.
(The Brownian regime at large $\epsilon$ is correlated in time at the scale of the 
Stokes time $\tau_S$, and one expects particle pairs with a relative velocity that
is not too small, to
behave ballistically at separation $r<\rv$). 

At least in 1D, the predictions of perturbation 
theory begin to be valid only for rather large values of $\epsilon$.
The fact is that
clustering at  $\epsilon\lesssim 1$, and residual concentration fluctuations 
at $\epsilon\gg 1$, are very different in nature.
At small $\epsilon$, in first approximation,
the particle separation $\r$, for $r\ll\rv$ evolves as a diffusion 
process with diffusivity $\propto\epsilon r^2$. 
Clustering
arises therefore from trapping at $r=0$ of the particle pairs.
For large $\epsilon$, most particle pairs at $r<\rv$ behave 
instead ballistically. In this regime, concentration fluctuations 
and associate particle velocity modifications,
can be seen as corrections to ballistic motion at scale $\rv$.
They are at most only the remnants 
of the trapping behaviors that dominate the pair dynamics 
at $\epsilon\lesssim 1$.


\appendix
\setcounter{equation}{0}
\setcounter{section}{1}
\renewcommand{\theequation}{\thesection\arabic{equation}}
\section*{Appendix A. $\ $Calculation of $\rho^\smaldu(\r)$.}
The generalization of Eq. (\ref{r1delta}) to the calculation of
$\langle\r^\smaldu|\r\rangle$ and $\langle\r^\smalun\r^\smalun|\r\rangle$ is obvious.
The resulting integrals in the form
$$
\int\d t_1\d t_2\ldots\Big\langle F[\bxi;t_1,t_2,\ldots)
\xi(t_1)\xi(t_2)\ldots\Big\rangle
$$
are simplified
by repeated application of the functional derivation by part
formula $\langle F[\bxi]\xi_\alpha(t)\rangle=\langle\delta F[\bxi]/\delta\xi_\alpha(t)\rangle$,
with the relations
$\delta\xi_\alpha(t)/\delta\xi_\beta(t')$ $=\delta_{\alpha\beta}\delta(t-t')$
and $\delta r^\smalze_\alpha(t)/\delta\xi_\beta(t')=
q\delta_{\alpha\beta}\theta(t-t')(1-\ex^{t'-t})$ with $\theta(t)$ the Heaviside
step function [$\theta(t>0)=1$; $\theta(t<0)=0$]. After some algebra, we obtain the
result
\beq
\begin{array}{ll}
\Omega\rho^\smaldu(\r)
=(q^4/4)
\\
\times
\int_{-\infty}^0\d t_a\int_{-\infty}^{t_a}\d t_b (1-\ex^{t_a})^4
\ex^{2(t_b-t_a)}
\\
\times\partial_\alpha\partial_\beta\partial_\gamma\partial_\phi
\int\d^3 r_a\int\d^3 r_bg_{\alpha\beta}(\r_a)g_{\gamma\phi}(\r_b)
\\
\times
\rho[\r^\smalze(t_a)=\r_a,\r^\smalze(t_b)=\r_b|\r^\smalze(0)=\r].
\end{array}
\label{intermediate3}
\eeq
As in the 1D case, we expect that the time integrals be dominated by $|t|,|t'|\ll 1$
($\tau_S=1$),
so that $(1-\ex^{t_1})^4\ex^{2(t_2-_1)}\to t_1^4$ and the correlation matrix entering the
Gaussian joint PDF
$\rho[\r^\smalze(t_a)=\r_a,\r^\smalze(t_b)=\r_b|\r^\smalze(0)=\r]$ read:
\beq
\begin{array}{ll}
\langle[r^\smalze_\alpha(t_a)-r_\alpha][r^\smalze_\beta(t_b)-r_\beta]\rangle
\\
=(q^2/2)\delta_{\alpha\beta}(1-|t_a-t_b|/2)\,t_at_b.
\nonumber
\end{array}
\eeq
Exploiting isotropy and incompressibility, it is convenient to write 
the random velocity correlation in the form:
\beq
g_{\alpha\beta}(\r)=
\rv^2[\partial_\alpha\partial_\beta-\delta_{\alpha\beta}\nabla^2]C(r/\rv).
\label{C(x)}
\eeq
We see that $C(x)\delta_{\alpha\beta}$ is the spatial correlation for the vector potential 
for the field $\u$ and the Fourier transform $C_k=\int\d^3x$ $\ex^{-\i\k\cdot\x}C(x)$
is thus positive defined.
To calculate the concentration fluctuation variance, we set $r=0$. Writing in terms
of Fourier components and using Eq (\ref{C(x)}), we can then rewrite
the right hand side (RHS) of
Eq. (\ref{intermediate3}) in the following form:
$$
\begin{array}{ll}
\frac{q^4\rv^{10}}{4}
&\int_{-\infty}^0t_a^4\d t_a\int_{-\infty}^{t_a}\d t_b
\int\frac{d^3 k_a}{(2\pi)^3}\frac{\d^3 k_b}{(2\pi)^3}
C_{k_a\rv}C_{k_b\rv}
\\
&\times
[(\k_a\cdot\k_b)^2-(k_ak_b)^2]^2
Z_{t_at_b}
(\k_a,\k_b),
\end{array}
$$
and
\beq
\begin{array}{ll}
Z_{t_at_b}(\k_a,\k_b)&=\exp\Big[-\frac{q^2}{4}\Big((t_ak_a)^2+(t_bk_b)^2
\\
&+2t_at_b
(1-\frac{1}{2}(t_a-t_b))\k_a\cdot\k_b\Big)\Big]
\end{array}
\nonumber
\eeq
is the generating function for $\r^\smalze(t_{a,b})$ conditioned to
$\r^\smalze(0)=0$. The multiple integrals in $\rho^\smaldu$ are simplified passing to polar
coordinates:
$
(\rv k_1,\rv k_2)\equiv (\tilde k_1,\tilde k_2)=(s^{1/2}\cos\theta,s^{1/2}\sin\theta);
$
$
(\epsilon^{1/2}t_1,\epsilon^{1/2}t_2)=(\tau^{1/2}\cos\varphi,\tau^{1/2}\sin\varphi)
$
and
$
\k_1\cdot\k_2=k_1k_2z.
$
We are going to verify that the integrals in $\rho^\smaldu$
are dominated by $s,\tau\sim 1$ so that
the term $\frac{1}{2}(t_a-t_b))\k_a\cdot\k_b$ in $Z_{t_at_b}(\k_a,\k_b)$ can
be disregarded. In this case, the $\tau$ integral can be carried out explicitly, and we obtain
the result
\beq
\begin{array}{ll}
f(0)=&\frac{1}{64\pi^4\epsilon}\int_0^\infty \d s\int_0^{\pi/2}\d\theta
\int_{-1}^1\d z\int_0^{\pi/4}\d\varphi
\\
\\
&\frac{s^3(1-z^2)^2\sin^62\theta\cos^4\varphi\
C_{\tilde k_a}C_{\tilde k_b}
}{[1+\cos(2(\theta+\varphi))
+(1+z)\sin 2\theta\sin 2\varphi]^3}.
\end{array}
\label{f(0)3D}
\eeq
From Eqs. (\ref{eq1},\ref{C(x)}), we have that $C_{\tilde k\lesssim 1}\sim 1$, so that,
provided the integrals in the RHS of Eq. (\ref{f(0)3D}) converge, we have
$f(0)\propto\epsilon^{-1}$. Convergence is also sufficient to verify correctness
of the ansatz $s,\tau\sim 1$ in the integral.
It is sufficient to prove convergence near the singularity at
$\theta=\pi/2$, $\phi=0$, $z=-1$, where the integrand is
$\propto\tilde\theta^6\tilde z^2\ (\tilde\varphi^2$ $+\tilde z\tilde\theta^2)^{-3}$
with $\tilde\varphi=\varphi+\theta-\pi/2$, $\tilde\theta=\pi/2-\theta$ and $\tilde z=1+z$.
Integrating first in $\d\tilde\varphi$, leads to an expression whose leading
term in $\tilde z$ and $\tilde\theta$ is $\propto\tilde\theta\tilde z^{-1/2}$,
the remaining integrals converge as required, and therefore
$\rho^\smaldu=O(\epsilon^{-1})$ as expected.

\setcounter{equation}{0}
\setcounter{section}{2}
\section*{Appendix B. $\ $ Collision velocity variance}

\subsection{The 1D case}
Substituting Eqs. (\ref{order0v}) and  (\ref{order1}) into
Eq. (\ref{nu21D}), we easily obtain:
\beq
\begin{array}{ll}
\langle\nu^2&|r\rangle^\smalun_{cl}=
\frac{q^3}{2}\int_\infty^0\d\tau_1\int_\infty^0\d\tau_2\int_\infty^0\d\tau_3
(1
\\
&-\ex^{\tau_1})\ex^{\tau_2+\tau_3}
\Big\langle\partial_{r^\smalze}g(1)\xi(1)\xi(2)\xi(3)
\\
&\times\delta(r^\smalze(0)-r)\Big\rangle_{2\ne 3}
\end{array}
\label{intermediate5}
\eeq
where $g(k)\equiv g(r^\smalze(\tau_k))$ and $\xi(k)\equiv\xi(\tau_k)$, $k=1,2,3$.
The subscript $2\ne 3$ at the end of the integrand indicates that the contraction
$\xi(2)\xi(3)\to \delta(\tau_2-\tau_3)$ is canceled by the identical contribution
coming from the average $\langle\xi(\tau_2)\xi(\tau_3)\rangle$ entering the
$\langle(\nu^\smalze)^2\rangle$ in the last term of Eq. (\ref{order0v}).
Repeated application of the functional integration by part formula leads
quickly to the following expression for the RHS of Eq. (\ref{intermediate5}):
$$
\begin{array}{ll}
-\int_{-\infty}^0\d\tau (1-\ex^\tau)^2
\Big[\frac{q^4}{2}\ex^\tau\partial_r^2
\\
+\frac{q^6}{8}(1-\ex^\tau)^2\partial_r^4\Big]
\Big\langle g(\r^\smalze(\tau))\delta(r^\smalze(0)-r)\Big\rangle.
\end{array}
$$
The calculation follows the same lines of those leading to Eq. (\ref{f(r)}).
The averages can be calculated using Eq. (\ref{sigma2s}) for $s(\tau)=r^\smalze(\tau)-r$.
It is convenient to express the result in terms of Fourier components; for $r=0$:
$$
\begin{array}{ll}
\frac{q^2}{2}\int\frac{\d k}{2\pi}g_k\int_{-\infty}^0\d\tau
\Big\{[qk(1-\ex^\tau)]^2\ex^\tau
\\
-\frac{1}{4}[qk(1-\ex^\tau)]^4\Big\}
\exp\Big\{-\frac{(qk)^2}{2}(-1-\tau+\ex^\tau)\Big\}
\end{array}
$$
As in the previous 1D calculations we find a logarithmic divergence, this time 
at $k\sim q^{-1}$ and the leading contribution to the integral is
$$
\frac{q^2g_0}{2\pi}\int_{q^{-1}}^\infty\d k
\Big[(qk\tau)^2-\frac{1}{4}(qk\tau)^4\Big]\exp\Big\{-\frac{(qk\tau)^2}{4}\Big\}.
$$
Using  $g_0=\int\d xg(x)=2\rv$ and $\langle(\nu^\smalze)^2\rangle=q^2/2$, we 
obtain the final result, that is Eq. (\ref{nu21Dfinal}):
\beq
\langle(\nu^\smalze)^2\rangle^{-1}\langle\nu^2|0\rangle_{cl}^\smalun
=-4(\pi\epsilon)^{-1/2}\ln cS,
\label{finalA}
\eeq
with $c$ an  $O(1)$ constant.

\subsection{The 3D case}
Calculation of $\langle\nu^2|0\rangle_{cl}$ in 3D is only slightly more involved.
Using Eqs. (\ref{order0}-\ref{order2}) and (\ref{order0v}), the
RHS of Eq.  (\ref{nu23D}) can be expressed as the sum of three integrals:
\beq
\begin{array}{ll}
I_1=\frac{q^4}{8}\int_{-\infty}^0\d\tau_1\int_{-\infty}^0\d\tau_3\int_{-\infty}^0\d\tau_4\,
P
\\
\times\Big\langle\partial_\beta g_{\alpha\gamma}(1)\partial_\alpha g_{\beta\gamma}(1)\xi_\omega(3)
\xi_\omega(4)\delta(\r(0)-\r)\Big\rangle_{3\ne 4}
\end{array}
\label{I_1}
\eeq
\beq
\begin{array}{ll}
I_2=\frac{q^4}{4}\int_{-\infty}^0\d\tau_1
\int_{-\infty}^{\tau_1}\d\tau_2
\int_{-\infty}^0\d\tau_3\int_{-\infty}^0\d\tau_4\,
P
\\
\times(1-\ex^{\tau_1})\ex^{-\tau_1+\tau_2}
\Big\langle\partial_\beta g_{\alpha\gamma}(1)\partial_\alpha g_{\beta\phi}(2)
\\
\times\xi_\gamma(1)\xi_\phi(2)\xi_\omega(3)\xi_\omega(4)\delta(\r(0)-\r)\Big\rangle_{3\ne 4}
\end{array}
\label{I_2}
\eeq
\beq
\begin{array}{ll}
I_3=\frac{q^3}{8}\int_{-\infty}^0\d\tau_1\int_{-\infty}^0\d\tau_3\int_{-\infty}^0\d\tau_4\,
P
\Big\langle g_{\beta\phi}(1)
\\
\times
\partial_\beta g_{\phi\gamma}(1)\xi_\gamma(1)
\xi_\omega(3)\xi_\omega(4)\delta(\r(0)-\r)\Big\rangle_{3\ne 4}
\label{I_3}
\end{array}
\eeq
where $P\equiv P(\{\tau_k\})=(1-\ex^{\tau_1})\ex^{\tau_3+\tau_4}$. As in the 1D case, the
contractions between $\xi(3)$ and $\xi(4)$ are canceled by
the last normalization term in Eq. (\ref{nu23D}). 

In analogy with the analysis of Eq. (\ref{small}), the magnitude of the three 
integrals $I_{2,2,3}$ can be estimated
exploiting the fact that the integrals
are dominated by $\tau_{1,2}\sim\epsilon^{-1/2}\ll 1$ because of  the factors 
$g(r^\smalze(\tau_{1,2})$, and using everywhere 
$\tau\sim\epsilon^{-1/2}$,
(then also $P\sim\tau$) and 
$\partial_r\sim\rv^{-1}$. 
Let us pass to the noise terms and consider first $\xi_\omega(3,4)$. 
The contractions with $\xi(1,2)$ lead to $\int\d\tau_k\ex^{\tau_k}\xi(3,4)\to 1$.
The same result is produced from action of $\xi(3,4)$ on
$g(1)g(1,2)\delta(\r^\smalze(0)$ $-\r)$, that is evaluated with the
functional integration by part formula. We obtain in fact:
$
\int\d\tau_k\ex^{\tau_k}$ $\xi_\omega(k)\to q(1-\ex^{\tau_{1,2}})\partial_\omega
\sim -q\tau_{1,2}\rv^{-1}\sim 1
$
In the same way we find:
$
\xi_\gamma(1)\to q(1-\ex^{\tau_2})\partial_\gamma\sim q\tau_2\rv\sim 1,
$
and $\xi_\phi(2)\sim 1$. 
Substituting into Eqs. (\ref{I_1}-\ref{I_3}), we find $I_{1,3}=O(q^2\epsilon^{-1/2})$
and $I_2=O(q^2\epsilon^{-1})$. We thus obtain to $O(q^2\epsilon^{-1})$:
$$
\begin{array}{ll}
\langle\nu^2|\r=0\rangle=-\frac{q^2}{8}\int_{\infty}^0\tau^2\d\tau
\Big\langle\partial_\alpha g_{\beta\gamma}(\r^\smalze(\tau))
\\
\qquad\qquad\quad\times\partial_\beta g_{\alpha\gamma}(\r^\smalze(\tau))\delta(\r^\smalze(0))\Big\rangle
\\
=-\frac{q^2}{8}\int_{\infty}^0\tau^2\d\tau\int\d^3r_a\ 
\rho[\r^\smalze(\tau)=\r_a|\r^\smalze(0)=0]\,
\\
\qquad\qquad\qquad\qquad\quad\times\partial_\alpha g_{\beta\gamma}(\r_a)
\partial_\beta g_{\alpha\gamma}(\r_a)
\end{array}
$$
Passing to Fourier components, and using Eq. (\ref{C(x)}) we obtain the final 
result:
\beq
\begin{array}{ll}
\frac{\langle\nu^2|\r=0\rangle^\smaldu}
{\langle(\nu^\smalze)^2\rangle}
=
\frac{-1}{24\epsilon^{1/2}}
\int_{-\infty}^0\tau^2\d\tau\int\frac{\d^3k}{(2\pi)^3}\frac{\d^3l}{(2\pi)^3}
C_k
C_l
\\
\qquad\qquad
\times 
k^3l^3(1-z^2)z\,
\exp(-|\k+\l|^2\tau^2/4),
\end{array}
\label{finalB}
\eeq
where $z=(\k\cdot\l)/(kl)$,
$\langle(\nu^\smalze)^2\rangle=3q^2/2$
and $\exp(-|\k+\l|^2\tau^2/4)=Z_{\epsilon^{1/2}\tau}(\rv(\k+\l)$, 
with $Z_t(\k)$ the generating function for $\r^\smalze(\tau)$ conditioned to 
$\r^\smalze(0)=0$.

The angular integral in Eq. (\ref{finalB}), is in the form
$\int_{-1}^1\d z\,(1-z^2)z\exp(-(k^2+l^2+2klz)\tau^2/4)<0$,
and this tells us that the constant $H$ in Eq. (\ref{nu23Dfinal}) is positive.

\end{document}